# Ultra-thin Underwater Acoustic Metasurface with Multiply Resonant Units

Haozhen Zou, Pan Li, and Pai Peng*

School of Mathematics and Physics, China University of Geosciences, Wuhan 430074, China

**Abstract**

This paper describes a new kind of acoustic metasurface with multiply resonant units, which have previously been used to induce multiple resonances and effectively produce negative mass density and bulk/shear moduli. The proposed acoustic metasurface can be constructed using real materials and does not rely on an ideal rigid material. Therefore, it can work well in a water background. The thickness of the acoustic metasurface is about two orders of magnitude smaller than the acoustic wavelength in water. The design of a unit group is proposed to avoid the phase discretization becoming too fine in such a long-wavelength condition. We demonstrate that the proposed acoustic metasurface achieves good performance in anomalous reflection, focusing, and carpet cloaking.

*Corresponding author: paipeng@cug.edu.cn

# Introduction

Acoustic metasurfaces (AMs)[1,2] that offer phase modulation have attracted a lot of interests in recent years. Generally, AMs can be constructed using one layer of resonators, which can provide large phase delays relative to the matrix around the resonant frequencies. Several kinds of resonators have been wildly used and systematically studied in elastic metamaterials. The Helmholtz resonators[3] (or bubbles in water) are used to induce monopolar resonance, giving the elastic metamaterial an effectively negative bulk modulus. The three-component resonators[4,5] (or membranes[6]) are used to induce dipolar resonance, resulting in an effectively negative mass density. The multiple resonators[7] are used to induce strong quadrupolar resonances, which brings about an effectively negative shear modulus. These three types of resonance (monopolar, dipolar, and quadrupolar) complete the set of negative parameters for elastic metamaterials. The first two kinds of resonators, the Helmholtz resonators[8,9] and the membranes[10] have been used to construct AMs. So far, there is no counterpart for AMs made by the multiple resonators reported. Here we present for the first time the AMs by using the multiply resonant units. Be different with those AMs with Helmholtz resonators or membranes, which rely on air background because they are made by, more or less, ideal rigid materials, the proposed AMs with multiply resonant units can be constructed using real materials and show good performances in a water background.

In this paper, the multiply resonant unit we use are the same as that introduced in Ref.6. The resonator contains a multi-mass resonant unit and can induce quadrupolar resonance, as well as monopolar and dipolar resonance. The multiply resonant units induce a large bandgap for dipolar resonance and two negative bands (inside the bandgap), which are caused by double-negative metamaterials[11, 12], indicated the overlap regions between the monopolar or quadrupolar resonance and the dipolar resonance. Here, we focus on the frequencies in the second negative band, which is induced by simultaneous monopolar and dipolar resonance.

**Anomalous Acoustic Reflection**

We consider the unit cell shown in Fig. 1(a), which consists of a rectangular foam host and an inclusion. As shown in Fig. 1(b), the inclusion is composed of a soft rubber cylinder embedded with a hard rubber cylinder, surrounded by four rectangular steel rods in the shape of an X. The tilt angle of the steel rods is 45°. The material parameters are as follows: $\rho_f = 115 \text{kg/m}^3$, $\kappa_f = 9 \times 10^6 \text{N/m}^2$, $\mu_f = 3 \times 10^6 \text{N/m}^2$ for the foam; $\rho_{sr} = 1.3 \times 10^3 \text{kg/m}^3$, $\kappa_{sr} = 6.4 \times 10^5 \text{N/m}^2$, $\mu_{sr} = 4 \times 10^4 \text{N/m}^2$ for the soft rubber; $\rho_{hr} = 1.415 \times 10^3 \text{kg/m}^3$, $\kappa_{hr} = 1.27 \times 10^9 \text{N/m}^2$, $\mu_{hr} = 1.78 \times 10^6 \text{N/m}^2$ for the hard rubber; and $\rho_s = 7.9 \times 10^3 \text{kg/m}^3$, $\kappa_s = 1.94 \times 10^{11} \text{N/m}^2$, $\mu_s = 8.28 \times 10^{10} \text{N/m}^2$ for the steel, where $\rho$, $\kappa$, and $\mu$ denote the mass density, bulk, and shear modulus, respectively. The geometric parameters are $h$ and $2h$ for the host width and length, respectively; $r_1 = 0.4h$ and $r_2$ for the radii of the soft and hard cylinders, respectively; $a = 0.24h$ and $b = 0.16h$ for the length and width of the steel rods, respectively. The radius $r_2$ is used to adjust the resonant frequency of the unit. The distance from the center of the steel rods to the center of the rubber cylinder is $0.24h$, and the inclusion is placed

$0.5h$ from the right-hand side of the foam. As shown in Fig. 1(a), the resonant units are periodically arranged in the water background, and ideal rigid barriers are set on the right-hand side and between the units to ensure that the waves are totally reflected and that the neighboring units are independent of each other, respectively.

Plane acoustic waves with angular frequency $\omega = 2\pi c / \lambda$ are generated and normally incident from the left, and these are totally reflected by the units, where $c = 1490 \text{m/s}$ is the speed of sound in water and $\lambda = 70h$ is the corresponding wavelength. We use a finite element method within the commercial COMSOL Multiphysics software to calculate the phase for the reflected waves. The results are plotted in Fig. 1(c), where the reflected phase $\varphi$ is a function of the radius $r_2$ at the working frequency. The black solid line shows the phase delay of a full $2\pi$ span, which is uniformly discretized into 10 parts and realized by 10 units with different radii $r_2$, as shown by the red dots in Fig. 1(c). The corresponding reflected pressure fields for the units are shown on the left of Fig. 1(c), and the phase difference between the units has a step size of $\pi / 5$.

The proposed units can be used to build AMs. Note that the unit width is one-seventieth of the wavelength in water. To prevent the phase gradient from becoming too small along the AM, we design the unit groups as sub-structures. As shown in Fig. 2(b), the AM consists of two supercells. Each supercell is composed of 10 unit groups, which are further composed of 14 identical multiply resonant units, as shown in Fig. 2(c). The units in neighboring groups have a phase difference of $\pi / 5$. In accordance with the generalized Snell's law[13], the additional wave vector (phase gradient) along the metasurface direction can be defined as $\xi = d\varphi/dy$, where $d\varphi$ is the discretized phase step and $dy$ is the distance between two units with that phase step. If there is no unit group, then $d\varphi = \pi/70$ and $dy = h$. Obviously, $d\varphi$ is too small and will introduce unnecessary complexity. Thus, we use a coarser discretization of the phase gradient via the unit group. Then, for the unit group, we have $d\varphi = \pi/5$ and $dy = 14h$. When the working frequency is very low, corresponding to a very long wavelength compared to the inside geometry, the unit group is particularly useful for simplifying. The additional wave vector $\xi$ is not changed by the unit group. The properties of reflected waves can be obtained by employing the continuity boundary condition [16] in the form $k\sin\theta_r = \xi + k\sin\theta_i$, as shown in the illustration of Fig. 2(a). Here, $k = 2\pi/\lambda$ is the wave vector in water and $\theta_i$, $\theta_r$ are the incident and reflected angles, respectively. In the case of normal incidence, we have $\theta_i = 0$ and the reflected angle $\theta_r = \arcsin(\xi/k)$, which in this case is $\theta_r = \arcsin(1/2) = 30°$. The pressure field of the reflected waves is plotted in Fig. 2(a), and an anomalous acoustic reflection can be observed. The dashed black lines show the theoretical prediction of the reflected angle, which is in good agreement with the simulations.

**Acoustic Focusing and Carpet Cloaking**

A design for a planar AM lens[14, 15] using the multiply resonant units is now presented, and the process of acoustic focusing is demonstrated. As shown in Fig. 3(a), to obtain constructive interference at the designed focal point, the reflected phases should be a function of position and satisfy the equation $\varphi(0) - \varphi(y) = kl(y)$, where $\varphi(0)$ and $l(y)$ are the reflected phase at the center of the planar lens and the relative acoustic path difference, respectively. The planar lens is realized by an AM consisting of 31 unit groups, each composed of 20 identical units. The phase profile $\varphi(y)$ of the AM along the $y$ direction is shown in Fig. 3(b), following the design in Fig. 1(a). The reflected phase at the center is set to $\varphi(0) \approx 2\pi$, and the designed focal length is $f = 5\lambda$. When acoustic waves are normally incident from the left, the reflected pressure field is shown in Fig. 3(c), where an acoustic focusing effect can be observed. The distribution of the amplitude of reflected pressure along the lens center (indicated by the black dashed line) is plotted in Fig. 3(d). The pressure $p(x, y = 0)$ from Fig. 3(c) is shown as a function of position $x$, after normalization according to its maximum value. We can see that the maximum pressure, which is the focal point obtained from the simulations, occurs at a distance of approximately $4.7\lambda$ from the AM lens. The simulation results of the focal length are in good agreement with the theoretical design, as shown by the red and blue arrows in Fig. 3(d). The corresponding distribution of the amplitude of reflected pressure along the red dashed line (at the focal length) is plotted in Fig. 3(e), where the pressure is a function of position $y$ given by $p(x = -4.7\lambda, y)$ and the curve has a half-peak bandwidth of approximately $\Delta y = 0.86\lambda$, and the planar AM lens exhibits good performance in terms of acoustic focusing.

The final case considered in this paper concerns acoustic carpet cloaking using the AM with multiply resonant units. As shown in Fig. 4(a), the inclusion inside the unit is rotated 45° and placed orthogonally. The inclusion is the same as that used previously, although this rotation means that the X shape is now a cross. The corresponding reflected phase $\varphi$ as a function of the radius $r_2$ at the same working frequency is shown by the black line in Fig. 4(a). As introduced in Ref.6, the multiply resonant units can induce simultaneous monopolar and dipolar resonance along different directions, although the resonator is anisotropic. Here, we demonstrate acoustic carpet cloaking[16, 17] using an AM with a cross-shaped inclusion.

As shown in Fig. 4(b), a rigid bump in the form of a triangular prism is placed on the right wall. The bump has a height of $w(0) = 0.48\lambda$ and base angles of $\beta = 14^0$. As shown in Fig. 4(b), when plane waves are normally incident from the left-hand side before being reflected by the bump, the reflected pressure field is messy. The AM carpet cloak is designed to eliminate the scattering from the bump. In the case of normal incidence, the phase difference caused by the bump should be $\varphi(y) = 2kw(y)$, where $w(y)$ is the distance from the bump surface to the right-hand wall. The distance is a function of position $y$, i.e., $w(y) = w(0) - y\tan\beta$. The proposed AM cloak consists of two supercells. Each supercell is composed of 10 unit groups, and each unit group is composed of 14 identical units. The phases for every unit group are designed to compensate the phase difference caused by the bump. The reflected pressure field in the presence of AM cloaking is shown in Fig. 4(c), where plane reflected waves are clearly observed. The reflected field is essentially the same as the field reflected by the wall without the bump. Therefore, the AM shows good performance in terms of cloaking the bump.

## Conclusion and Discussion

In conclusion, we have proposed a new kind of AM with multiply resonant units. The AM is constructed from real materials and does not rely on an ideal rigid material. Hence, the proposed AM works well in a water background. Although rigid barriers are set between the units, these are not strictly necessary and are only used to simplify the design. In real applications, removing the rigid barriers would not affect the performance of the AM, although the phase design would be more complicated. The thickness of the AM is about two orders of magnitude smaller than the wavelength in water. In the case of such long wavelength conditions, the AM can be homogenized. This would bring novel research perspectives. We have designed the unit groups so as to avoid the phase discretization becoming too fine. Finally, anomalous reflection, focusing, and carpet cloaking were demonstrated through numerical simulations.

## Acknowledgments

This work was supported by the National Natural Science Foundation of China (Grant No. 11604307).

## References

1. Y. Li, B. Liang, Z.-m. Gu, X.-y. Zou and J.-c. Cheng, Scientific reports **3**, 2546 (2013).

2. B. Assouar, B. Liang, Y. Wu, Y. Li, J.-C. Cheng and Y. Jing, Nature Reviews Materials **3**, 460 (2018).

3. N. Fang, D. Xi, J. Xu, M. Ambati, W. Srituravanich, C. Sun and X. Zhang, Nat Mater **5**, 452 (2006).

4. Z. Liu, X. Zhang, Y. Mao, Y. Zhu, Z. Yang, C. T. Chan and P. Sheng, Science **289**, 1734 (2000).

5. Z. Liu, C. T. Chan and P. Sheng, Physical Review B **71**, 014103 (2005).

6. Z. Yang, J. Mei, M. Yang, N. H. Chan and P. Sheng, Phys. Rev. Letter **101**, 204301 (2008).

7. Y. Lai, Y. Wu, P. Sheng and Z.-Q. Zhang, Nature materials **10**, 620 (2011).

8. Y. Li, X. Jiang, B. Liang, J.-c. Cheng and L. Zhang, Physical Review Applied **4**, 024003 (2015).

9. Y. Li, C. Shen, Y. Xie, J. Li, W. Wang, S. A. Cummer and Y. Jing, Physical review letters **119**, 035501 (2017).

10. G. Ma, M. Yang, S. Xiao, Z. Yang and P. Sheng, Nature materials **13**, 873 (2014).

11. J. Li and C. T. Chan, Physical Review E **70**, 055602 (2004).

12. Y. Wu, Y. Lai and Z.-Q. Zhang, Physical Review B **76**, 205313 (2007).

13. N. Yu, P. Genevet, M. A. Kats, F. Aieta, J.-P. Tetienne, F. Capasso and Z. Gaburro, Science **334**, 333 (2011).

14. W. Wang, Y. Xie, A. Konneker, B.-I. Popa and S. A. Cummer, Applied Physics Letters **105**, 101904 (2014).

15. B. Yuan, Y. Cheng and X. Liu, Applied Physics Express **8**, 027301 (2015).

16. C. Faure, O. Richoux, S. Félix and V. Pagneux, Applied Physics Letters **108**, 064103 (2016).

17. H. Esfahlani, S. Karkar, H. Lissek and J. R. Mosig, Physical Review B **94**, 014302 (2016).

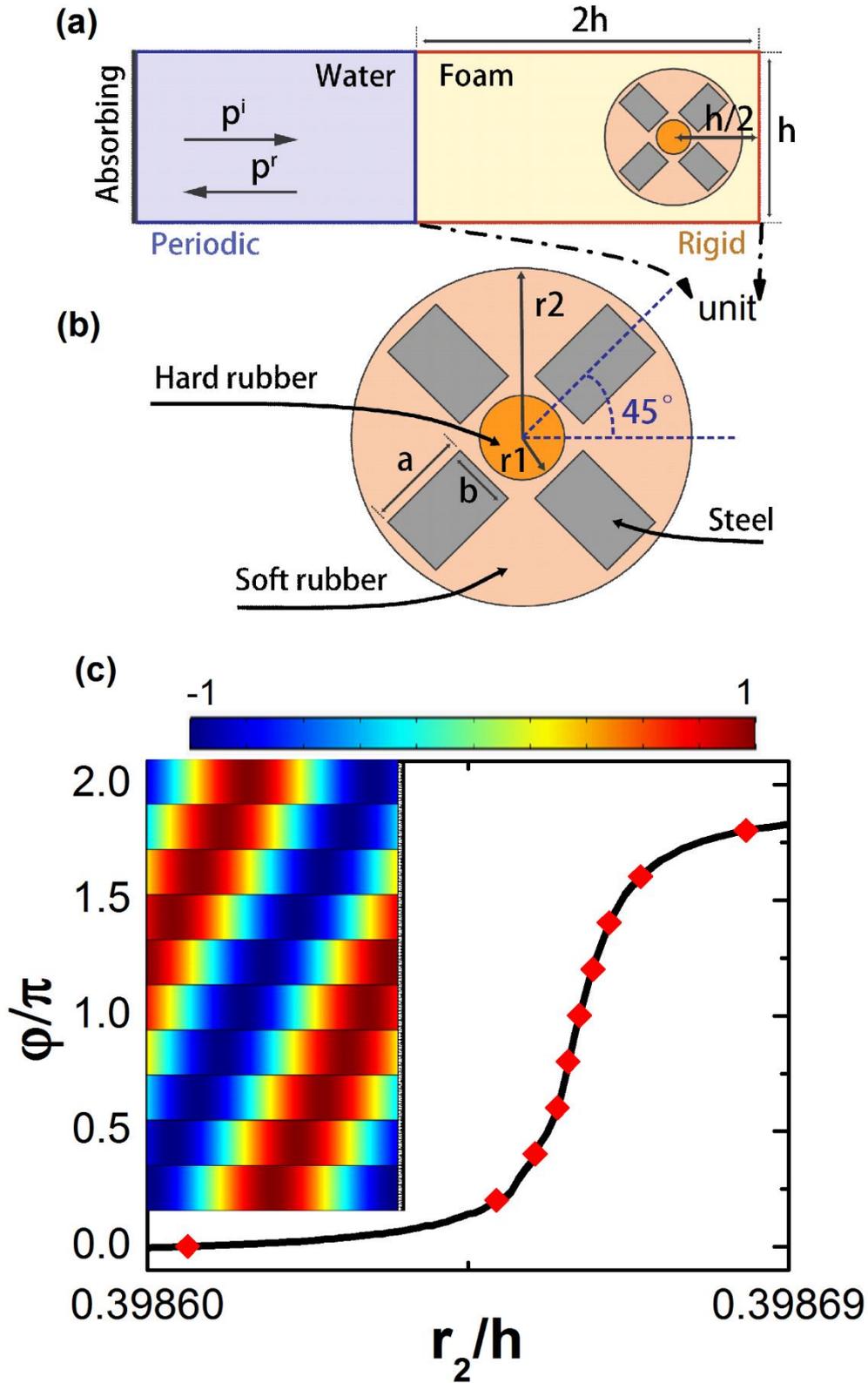

**FIG. 1** (a) is the calculation region. Rigid boundaries are set on the top, bottom, and right sides of the unit. Periodic boundaries are set on the top and bottom sides of the water, and an absorbing boundary is set on the left side. (b) shows a magnified view of the geometry of the inclusion. (c) is the reflected phase as a function of the radius $r_2$. The insets show the pressure fields of the reflected waves.

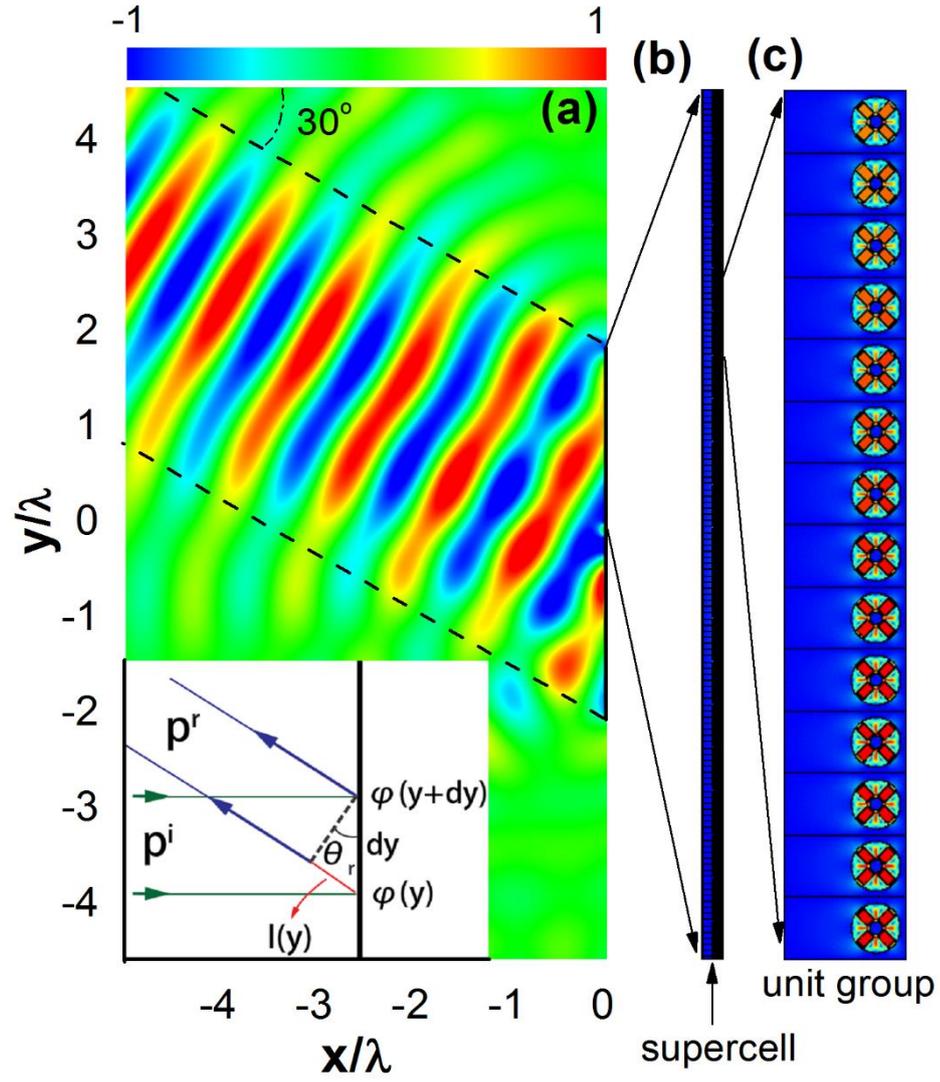

**FIG. 2** (a) is the reflected pressure field under an acoustic Gaussian beam normally incident from the left. The illustration is a schematic diagram of the anomalous reflection. (b) and (c) are schematic diagrams of a supercell and a unit group, respectively.

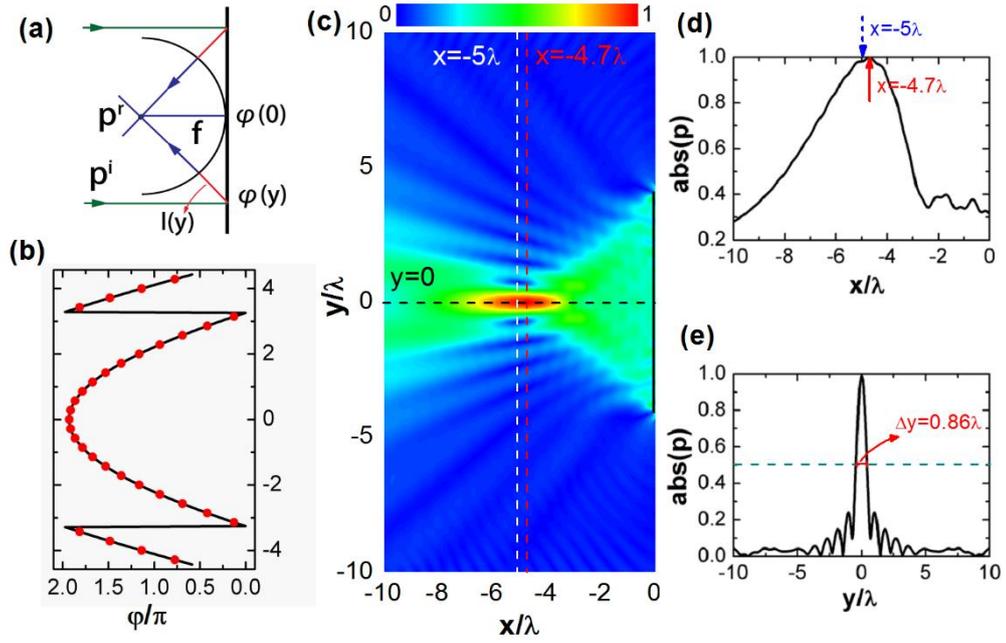

**FIG. 3** (a) Schematic diagram of a planar focusing lens based on a gradient AM. (b) is the phase profile realized by an AM lens. (c) is the reflected pressure field under an acoustic Gaussian beam normally incident from the left. The white and red dashed lines indicate the designed and calculated focal lengths, respectively. (d) is the reflected pressure normalized by the maximum value along the horizontal direction (at $y = 0$), and (e) is the corresponding pressure along the horizontal direction (at $x = -4.7\lambda$).

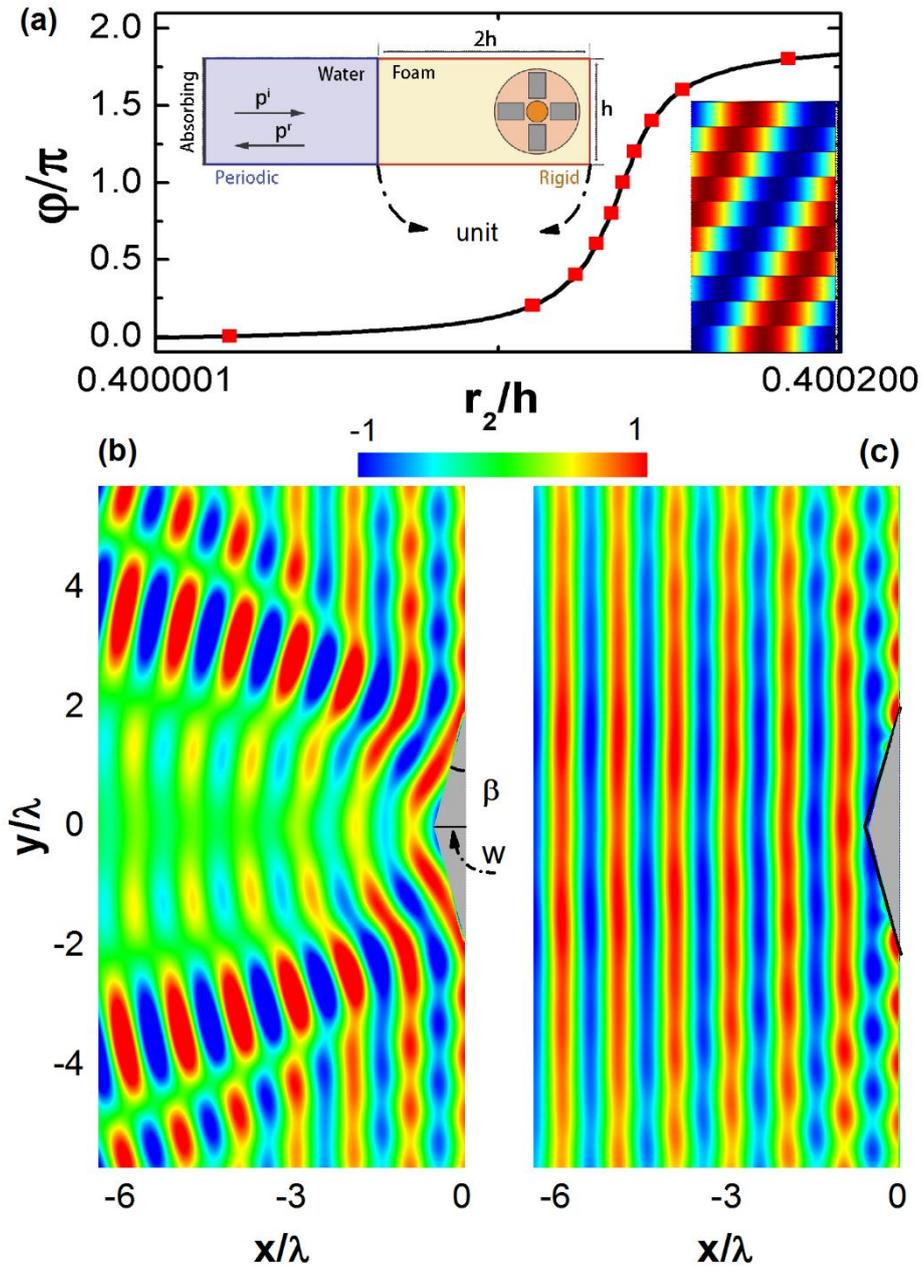

**FIG. 4** (a) is the reflected phase as a function of the radius $r_2$, where the inclusion is in the shape of a cross. (b) and (c) are the reflected pressure fields without and with the designed AM cloak, respectively. The object is a rigid bump, and the incident plane waves are coming from the left side.